\documentclass{svjour2}

\usepackage{graphicx}  
\usepackage{latexsym}
\usepackage{amssymb}
\usepackage{amsmath}

\def\beq{\begin{equation}}
\def\eeq{\end{equation}}

\journalname{General Relativity and Gravitation}

\begin{document}

\title{G\"odel spacetime, planar geodesics and the M\"obius map}

\author{Donato Bini \and 
Andrea Geralico\and 
Robert T. Jantzen\and 
Wolfango Plastino
}

\institute{Donato Bini
\at
Istituto per le Applicazioni del Calcolo ``M. Picone,'' CNR, Via dei Taurini 19, I-00185, Rome, Italy\\
INFN, Sezione di Roma Tre, I-00146 Rome, Italy\\
\email{donato.bini@gmail.com} 
  \and
Andrea Geralico 
\at
Istituto per le Applicazioni del Calcolo ``M. Picone,'' CNR, Via dei Taurini 19, I-00185, Rome, Italy\\
\email{andrea.geralico@gmail.com}
  \and
Robert T. Jantzen
\at
Department of Mathematics and Statistics, Villanova University, Villanova, PA 19085, USA\\
\email{robert.jantzen@villanova.edu}       
   \and
Wolfango Plastino
\at
Roma Tre University, Department of Mathematics and Physics, I-00146 Rome, Italy\\
INFN, Sezione di Roma Tre, I-00146 Rome, Italy\\
\email{wolfango.plastino@uniroma3.it }       
}

\date{Received: date / Accepted: date / Version: \today}

\maketitle

\begin{abstract}
Timelike geodesics on a hyperplane orthogonal to the symmetry axis of the G\"odel spacetime appear to be elliptic-like if standard coordinates naturally adapted to the cylindrical symmetry are used.
The orbit can then be suitably described through an eccentricity-semi-latus rectum parametrization, familiar from the Newtonian dynamics of a two-body system.
However, changing coordinates such planar geodesics all become explicitly circular, as exhibited by Kundt's form of the G\"odel metric.
We derive here a one-to-one correspondence between the constants of the motion along these geodesics as well as between the parameter spaces of elliptic-like versus circular geodesics.
We also show how to connect the two equivalent descriptions of particle motion by introducing a pair of complex coordinates in the 2-planes orthogonal to the symmetry axis, which brings the metric into a form which is invariant under M\"obius transformations preserving the symmetries of the orbit, i.e., taking circles to circles.
\end{abstract}

\section{Introduction}

The geometrical properties and physical aspects of G\"odel spacetime \cite{Godel:1949ga} which make it the prototype for rotating cosmological models have been investigated in depth in the literature since its discovery, leading to the introduction of several coordinate systems which are better suited for showing its various properties (see, e.g., Refs. \cite{kundt,chandra,Novello:1992hp,Obukhov:2000jf,Grave:2009zz,Bini:2014uua}).
Cylindrical-like coordinates are naturally adapted to the cylindrical symmetry of the spacetime about one particular dust particle world line \cite{Hawell}.
The dust source particles are at rest with respect to these coordinates, but form a family of twisting world lines, so that the cylindrical symmetry of the spacetime is preserved about every point.
Timelike geodesics on a ``planar" 2-surface orthogonal to the symmetry axis appear to be closed elliptic-like curves, which are of different types depending on the value of the ratio between the particle's conserved angular momentum ($L$) and energy $(E)$ \cite{Novello:1982nc}.
We have recently introduced an eccentricity-semi-latus rectum ($e,p$) parametrization to describe the motion in close analogy with the Newtonian dynamics of a two-body system \cite{Bini:2019xpv}. For each class of orbits we have then computed gauge-invariant quantities, like the periastron advance and the precession angle of a test gyroscope, thus providing coordinate-independent information.
The shape of the orbit, however, is instead coordinate-dependent.
The Kundt form of the G\"odel metric \cite{kundt} makes the orbits of all such planar geodesics explicitly circular.

The goal of this article is to derive a one-to-one correspondence between the constants of the motion and parameter spaces of the planar geodesics in the two different forms of the G\"odel metric mentioned above, i.e., the standard cylindrical form and Kundt's form.
We will also show how to connect the two equivalent descriptions of particle motion by introducing a pair of complex coordinates in the 2-planes orthogonal to the symmetry axis, which brings the metric into a form which is invariant under M\"obius transformations preserving the shape of the orbit, i.e., taking circles to circles.

\section{Planar timelike geodesics in the G\"odel spacetime}

The G\"odel spacetime \cite{Godel:1949ga,Hawell} is a Petrov type D stationary axisymmetric solution of the nonvacuum Einstein equations with a negative cosmological
constant and matter in the form of dust, whose gravitational attraction is balanced by a global rotation.
The stress-energy tensor is then $T=\rho\, u\otimes u$, with constant energy density $\rho$ and unit 4-velocity $u$ of the fluid particles aligned with the time coordinate lines, whereas the cosmological constant has the value $\Lambda=-\omega^2=-4\pi\rho$.
G\"odel spacetime is spacetime-homogeneous with five Killing vector fields which allow an infinite number of spatially homogeneous slicing families, each valid within a given neighborhood of some central world line. Changing ``spatial" coordinate systems inevitably changes the time slicing family as well.

\subsection{Cylindrical-like standard form and elliptic-like curves}

Using standard coordinates $(t,r,\phi,z)$ naturally adapted to the cylindrical symmetry \cite{Hawell}, the G\"odel metric reads
\begin{eqnarray}
\label{met_godel}
d s^2&=&\frac{2}{\omega^2}\left[-d t^2 +d r^2 +\sinh^2 r(1-\sinh^2 r)\,d \phi^2 \right.
\nonumber\\
&&\qquad \left. +2\sqrt{2}\sinh^2r\, d t\,d \phi  +d z^2\right]
\nonumber\\
&=&\frac{2}{\omega^2}\left[-(d t -\sqrt{2}\sinh^2 r\,d\phi)^2 +d z^2 \right.
\nonumber\\
&&\qquad \left.  +d r^2 +\sinh^2 r(1+\sinh^2 r)\,d \phi^2\right]
\,,
\end{eqnarray}
where the latter form is the orthogonal decomposition adapted to the time coordinate lines imbedded in the dust.
The coordinates here are all dimensionless, while the scaling factor $\omega>0$ has the dimension of inverse length and describes the vorticity of the dust world lines which rotate in the increasing $\phi$ direction about the central world line of the coordinate grid. 

The horizon radius
\beq
r_h=\ln (1+\sqrt{2})\approx 0.88137\,, \qquad
\sinh r_h=1\,,\quad 
\cosh r_h=\sqrt{2}\,,
\eeq 
is defined by the condition $ g_{\phi\phi}=0$,  where the $\phi$ coordinate circles are null and beyond which they are timelike. This radius delimits the ``physical region" of the spacetime for the given system of coordinates, if one wants to avoid closed timelike curves.

The Killing vectors associated with the spacetime symmetries are
\begin{eqnarray}
\label{killcyl}
\xi_1&=&\partial_t\,,\quad
\xi_2=\partial_\phi\,,\quad
\xi_3=\partial_z\,,\nonumber\\
\xi_4&=&\sqrt{2}\tanh(r)\cos\phi\,\partial_t-\sin\phi\,\partial_r-\frac{2\cos\phi}{\tanh(2r)}\,\partial_\phi\,,\nonumber\\
\xi_5&=&\sqrt{2}\tanh(r)\sin\phi\,\partial_t+\cos\phi\,\partial_r-\frac{2\sin\phi}{\tanh(2r)}\,\partial_\phi\,.
\end{eqnarray}
Introduce the 4-vector 
\beq
\eta=\sqrt{2}\tanh(r)\partial_t-\frac{2}{\tanh(2r)}\,\partial_\phi\,, \qquad
||\eta||^2=\frac1{\omega^2}(3-\cosh(4r))\,,
\eeq
which is spacelike in the region $0<r<r_*=\frac14{\rm arccosh}(3)\approx0.44069$ and timelike for $r_*<r<r_h$.
The Killing vectors $\xi_4$ and $\xi_5$ thus correspond to a rotation in the plane $r-\phi$
\beq
\begin{pmatrix}
-\xi_5\\
-\xi_4
\end{pmatrix}
=
\begin{pmatrix}
-\cos\phi&-\sin\phi\\
\sin\phi&-\cos\phi
\end{pmatrix}
\begin{pmatrix}
\partial_r\\
\eta
\end{pmatrix}
\,.
\eeq
Near the origin in the polar plane where $\eta\sim\partial_\phi$, this azimuthal rotation of the frame vectors $(\partial_r,\eta)$ aligns $(-\xi_5,-\xi_4)$ with the corresponding Cartesian frame $(\partial_x,\partial_y)$ defined in the usual way with respect to $(r,\phi)$, representing two translational Killing vector fields in that plane, though tilted with respect to the time coordinate hypersurfaces.
In fact when one spatially ``translates" the symmetry axis at the origin of these polar coordinates where the time slicing is orthogonal to the time lines, this tilting of these two  Killing vectors leads to a new slicing with the same property at the new location.

The geodesic equations are separable and the covariant 4-velocity of a general geodesic has the following  form
\beq
U^\flat=-E\,d t +L\,d \phi +b\,d z +U_r\, d r\,,
\eeq
where
\beq
U_r=\frac{2}{\omega^2}U^r=\frac{2}{\omega^2}\frac{dr}{d\lambda}\,,
\eeq
with $\lambda$ denoting an affine parameter and normalization condition $U^\alpha U_\alpha=-\mu^2$, with $\mu^2=1,0,-1$ for timelike, null, spacelike geodesics, respectively.
The geodesic equations expressed in terms of the constants of the motion
$E=-U\cdot \partial_t$, $L=U\cdot \partial_\phi$ and $b=U\cdot \partial_z$  are
\begin{eqnarray}
\frac{dt}{d\lambda}&=&-\frac12\omega^2\left(E-\frac{\sqrt{2}X}{\cosh^2r}\right)
\,,\nonumber\\
\left(\frac{dr}{d\lambda}\right)^2&=&-\frac12\omega^2\mu^2-\frac14\omega^4(b^2+E^2)\nonumber\\
&&+\frac14\omega^4\frac{X^2}{\cosh^2r}-\frac14\omega^4\frac{L^2}{\sinh^2r}
\,,\nonumber\\
\frac{d\phi}{d\lambda}&=&-\frac12\frac{\omega^2}{\sinh^2r}\left(\sqrt{2}E-\frac{X}{\cosh^2r}\right)
\,,\nonumber\\
\frac{dz}{d\lambda}&=&\frac12\omega^2b
\,,
\end{eqnarray}
where $X=\sqrt{2}E+L$. These can be integrated analytically.
Additional conserved quantities $k_4=U\cdot \xi_4$ and $k_5=U\cdot \xi_5$ associated with the Killing vectors $\xi_4$ and $\xi_5$ satisfy
\begin{eqnarray}
k_4\sin\phi+k_5\cos\phi&=&\frac{1}{\sinh(2r)}[\sqrt{2}E-(X+L)\cosh(2r)]\,,\nonumber\\
k_4\cos\phi-k_5\sin\phi&=&U^r\,,
\end{eqnarray}
which define  $r$ as an implicit function of $\phi$. 
This rotation of the Killing constants aligns the conserved translation Killing momenta with the polar frame momenta.

\subsubsection{Newtonian-like parametrization of planar geodesics}

Here we only consider the case $b=0$ of ``planar" timelike geodesics (choosing $\mu=1$ making  $\lambda=\tau$) confined to a constant $z$ hyperplane. 
The associated 4-velocity is
\begin{eqnarray}
   U &=& \frac{\omega^2}{2\cosh^2 r} \left[ \sqrt{2} L + E (2-\cosh^2 r) \right] \,\partial_t 
       +U^r \partial_r 
\nonumber\\
&& 
- \frac{\omega^2}{2 \sinh^2 r \cosh^2 r} \left( E \sqrt{2} \sinh^2 r - L\right) \,\partial_\phi
\,,
\end{eqnarray}
where 
\begin{eqnarray}
\label{Urequat}
(U^r)^2&=& \frac{\omega^2}{4\sinh^2 r \cosh^2 r}\left[-2(\tilde E^2+1)\cosh^4 r \right.\\
&+& \left. 2(3\tilde E^2+2\tilde E \tilde L+1)\cosh^2 r-(2\tilde E+\tilde L)^2  \right]
\,,\nonumber
\end{eqnarray}
having introduced the following rescaling of $E$ and $L$
\beq
\tilde E=\frac{\omega}{\sqrt{2}}  E\ge1\,, \qquad
\tilde L=\omega L\,.
\eeq
The orbits are elliptic-like and can be classified into three different types according to the ratio between angular momentum and energy \cite{Novello:1982nc}.

In Ref.~\cite{Bini:2019xpv} we have recently introduced an equivalent parametrization of the orbit in terms of a semi-latus rectum $p$ and eccentricity $e$, familiar from Newtonian mechanics (but for the change $r\to\cosh^2 r$), i.e.,
\beq
\label{rdichirel}
\cosh^2 r=\frac{p}{1+e\cos\chi}\,,
\eeq
with $p\geq1$ and $0\leq e<1$ as usual for elliptical orbits, so that $r=r_{\rm (peri)}$ for $\chi=0$ and $r=r_{\rm (apo)}$ for $\chi=\pi$, i.e., 
\beq
\label{rpmdef}
\cosh^2r_{\rm (peri)} = \frac{p}{1+ e}\,,\qquad
\cosh^2r_{\rm (apo)} = \frac{p}{1- e}\,.
\eeq
The main properties of the different kinds of orbits and the relation between the two parametrizations are summarized below.

\begin{figure*}
\begin{center}
$\begin{array}{cc}
\includegraphics[scale=0.3]{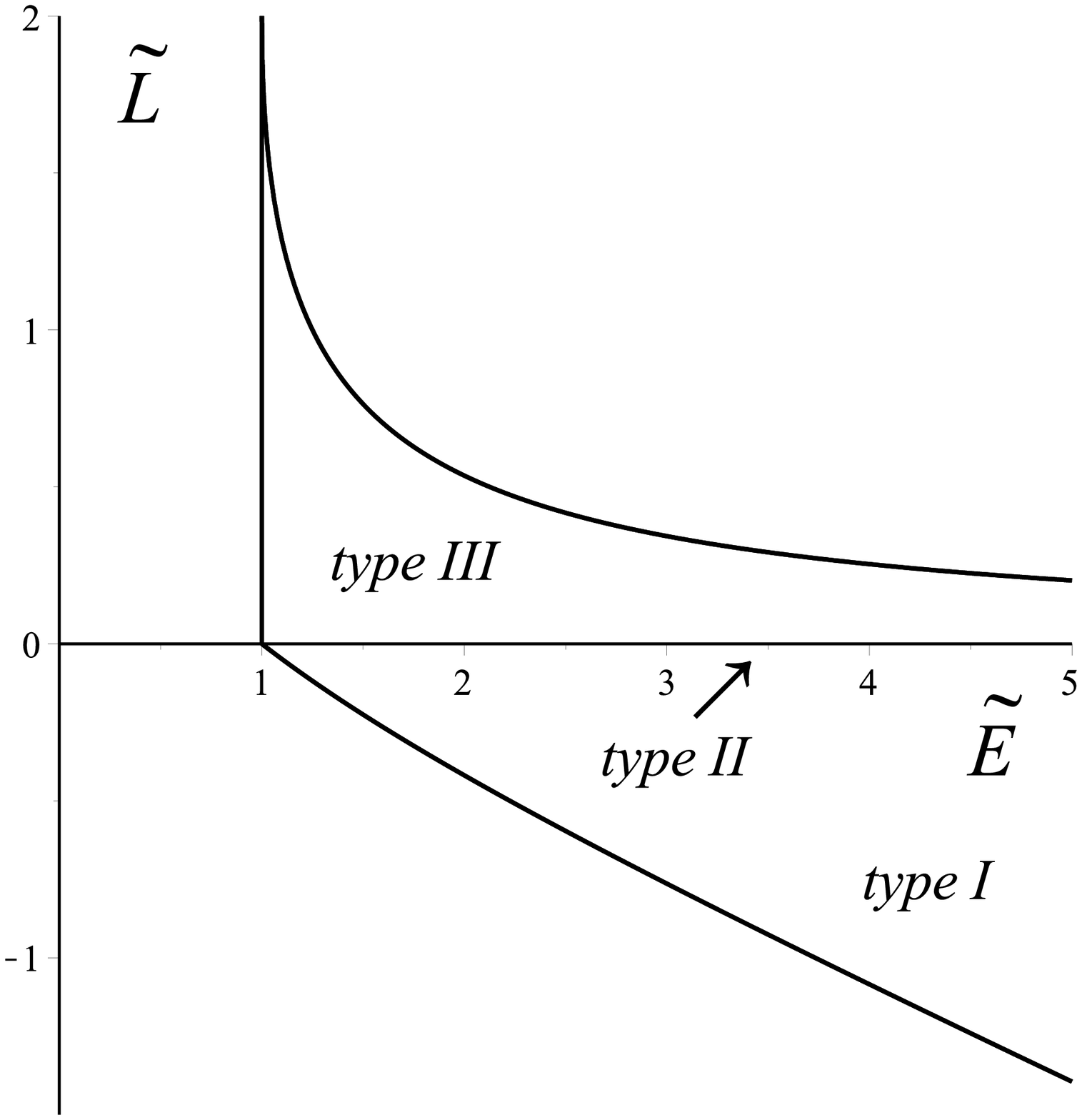}&\qquad
\includegraphics[scale=0.3]{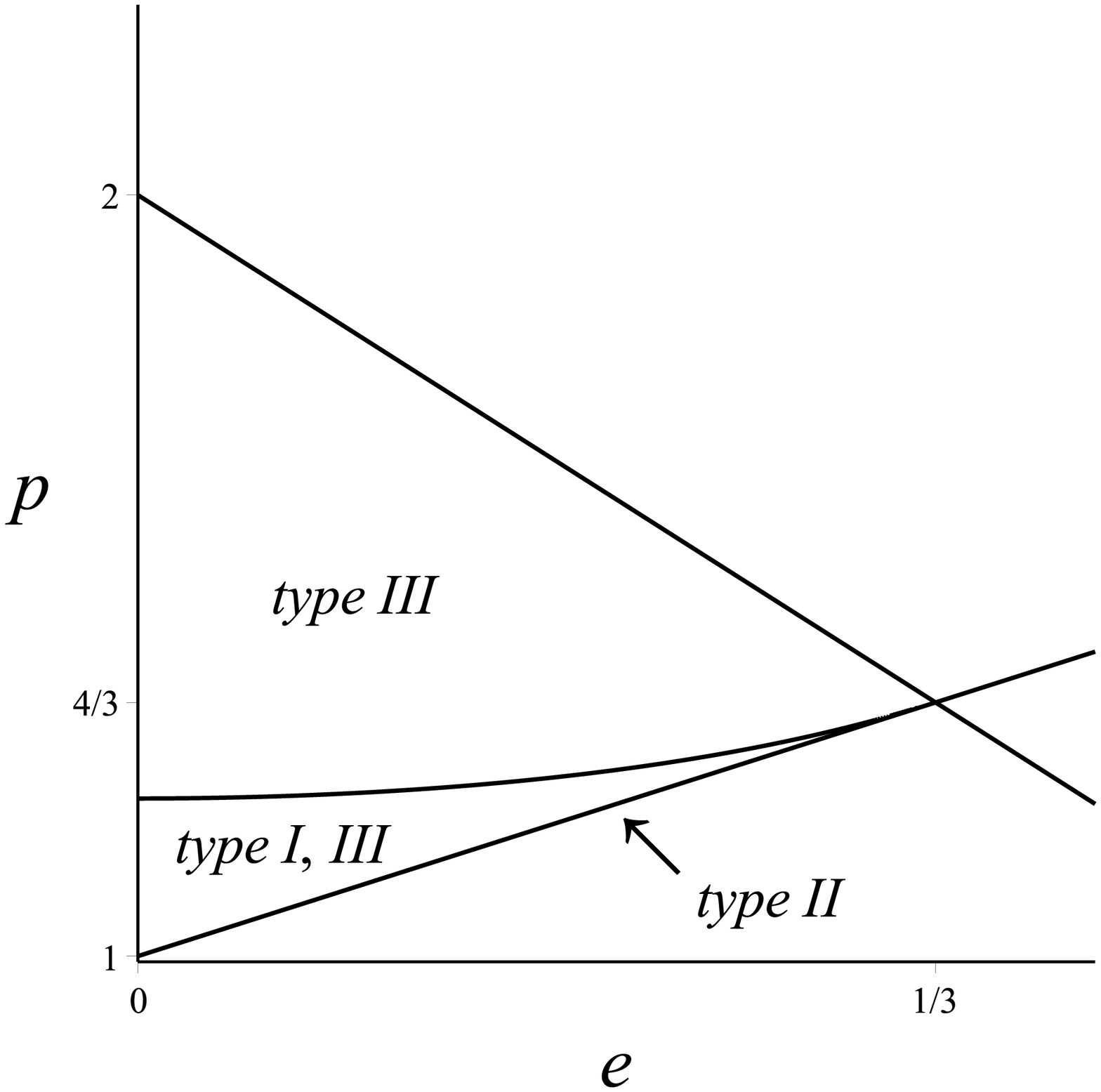}\\[.2cm]
\mbox{(a)} &\qquad \mbox{(b)}\cr
\end{array}
$\\
\end{center}
\caption{\label{fig:1} The allowed parameter spaces $(\tilde L,\tilde E)$ and $(p,e)$ for planar timelike geodesics in the physical region $r<r_h$ are shown in panels (a) and (b), respectively. 
Type II orbits have $\tilde L=0$, corresponding to the line $p=1+e$ where $r_{\rm (peri)}=0$ (see Eq.~(\ref{rpmdef})).
The type I ($\tilde L_{\rm min}\leq\tilde L<0$) and III ($0<\tilde L< L_h$) orbits are folded over onto each other into the triangular region between the lines $p=1+e$ and $p=2(1-e)$ where $r_{\rm (apo)}=r_h$. 
The upper boundary for type I orbits is implicitly defined by the vanishing of the denominator in Eq.~(\ref{tilde_E_L}), for which these geodesics become null.
}
\end{figure*}

\begin{enumerate}

\item type I:

The orbits have $0<r_{\rm (peri)}\leq r\leq r_{\rm (apo)}$ and go around the origin with monotonically decreasing $\phi$ opposing the local rotation of the dust source (no turning point in $\phi$).
The orbit becomes circular for $r_{\rm (peri)}=r_{\rm (apo)}$.
The angular momentum is always negative $\tilde L_{\rm min}\leq\tilde L<0$ for fixed values of the energy parameter, with 
\beq
\tilde L_{\rm min}=-\tilde E+\frac{1}{\sqrt{2}}\sqrt{\tilde E^2+1}\,.
\eeq
The relation between $(\tilde E,\tilde L)$ and $(p,e)$ is
\begin{eqnarray}
\label{tilde_E_L}
\tilde E &=& \frac{p+ {\mathcal Y}}{\sqrt{(p-{\mathcal Y})^2-4p(p-1)}}
\,,\nonumber\\
\tilde L &=& \frac{-2{\mathcal Y}}{\sqrt{(p-{\mathcal Y})^2-4p(p-1)}}\,,
\end{eqnarray}
where ${\mathcal Y}=\sqrt{(p-1)^2-e^2}$.

The solutions for $t$ and $\phi$ turn out to be 
\begin{eqnarray}
\label{phidichitypeI}
t(\chi)&=&\frac{1}{\sqrt{2}}\left[\chi-\frac{p+{\mathcal Y}}{\sqrt{1-e^2}} {\rm arctan} (\psi_1)\right]
\,,\nonumber\\
\phi(\chi) &=&-\frac{\chi}{2}-{\rm arctan} (\psi_2)\,,
\end{eqnarray}
with
\beq
\psi_1= \sqrt{\frac{1-e}{1+e}}\tan \frac{\chi}{2}\,,\qquad 
\psi_2= \sqrt{\frac{p-1+e}{p-1-e}} \tan \frac{\chi}{2}\,.
\eeq

\item type II:

The same as type I, with $r_{\rm (peri)}=0$, and $\cosh^2r_{\rm (apo)}=2\tilde E^2/(\tilde E^2+1)$, so that the orbits pass through the origin.
These orbits have vanishing angular momentum $\tilde L=0$ corresponding to $p=1+e$, with $\tilde E=\sqrt{{(1+e)}/{(1-3e)}}$.

The solutions for $t$ and $\phi$ turn out to be 
\begin{eqnarray}
\label{phidichitypeII}
t(\chi) &=&\frac{1}{\sqrt{2}}\left[\chi-\sqrt{\frac{1+e}{1-e}} {\rm arctan} (\psi_1)\right]
\,,\nonumber\\
\phi(\chi) &=&-\frac{\chi}{2}\,.
\end{eqnarray}

\item type III:

In this case the $\phi$-motion has a turning point in the allowed range of $r$, preventing the orbit from making a complete circuit about the origin.
The angular momentum is always positive $0<\tilde L< L_h$ for fixed values of the energy parameter, with 
\beq
L_h=2(\tilde E-\sqrt{\tilde E^2-1})\,.
\eeq
The relation between $(\tilde E,\tilde L)$ and $(p,e)$ is
\begin{eqnarray}
\tilde E &=& \frac{p-{\mathcal Y}}{\sqrt{(p+{\mathcal Y})^2-4p(p-1)}}
\,,\nonumber\\
\tilde L &=& \frac{2{\mathcal Y}}{\sqrt{(p+{\mathcal Y})^2-4p(p-1)}}\,.
\end{eqnarray}

The solutions for $t$ and $\phi$ turn out to be 
\begin{eqnarray}
\label{phidichitypeIII}
t(\chi)&=&\frac{1}{\sqrt{2}}\left[\chi+\frac{p-{\mathcal Y}}{\sqrt{1-e^2}} {\rm arctan} (\psi_1)\right]
\,,\nonumber\\
\phi(\chi) &=&-\frac{\chi}{2}+{\rm arctan} (\psi_2)\,.
\end{eqnarray}

\end{enumerate}
In any case the eccentricity is limited to $e<1/3$.
Furthermore, the relation between the variable $\chi$ and the proper time $\tau$ is 
\begin{eqnarray}
\label{chi_sol_fin}
 \tan \frac{\chi}{2}
&=&\sqrt{\frac{1+e}{1-e}}\tan \left(\zeta \tau  \right)\,, \quad \hbox{or}\nonumber\\
\tau &=& \frac{1}{\zeta}{\rm arctan}\left( \sqrt{\frac{1-e}{1+e}} \tan \frac{\chi}{2}\right)
\,,
\end{eqnarray}
with $\chi=0\leftrightarrow \tau=0$ and 
\beq
\label{zetadef}
\zeta =  \frac{ \omega\sqrt{\tilde E^2+1}}{  \sqrt{2}}\,.
\eeq
The allowed parameter spaces for both the $(\tilde L,\tilde E)$ and $(p,e)$ parametrizations are shown in Fig.~\ref{fig:1}.

Since the spacetime is homogeneous it is enough to consider geodesics of a given kind, e.g., type II orbits passing through the origin, without any loss of generality.
One can then switch to a geodesic of a different kind (type I and III) simply by changing the origin of the cylindrical-like coordinate system.
In fact, the classification of the orbits depends on the value of the angular momentum parameter $\tilde L$ relative to the origin $r =0$, so that it is always possible
to find a new parameter $\tilde L'$ with respect to the origin $r'=0$ of the new coordinate system such that old type II orbits will exhibit the same features as either type I or type III orbits.

The transformation $r\to r-r_0=r'$ implies that the new orbital parameters $(p',e')$ are related to the old ones $(p,e)$ by the conditions
\beq
\label{rppmdef1}
\cosh^2(r'{}_{\rm (peri)}+r_0) = \frac{p}{1+ e}\,,\qquad
\cosh^2(r'{}_{\rm (apo)}+r_0) = \frac{p}{1- e}\,,
\eeq
and
\beq
\label{rppmdef2}
\cosh^2r'{}_{\rm (peri)} = \frac{p'}{1+ e'}\,,\qquad
\cosh^2r'{}_{\rm (apo)} = \frac{p'}{1- e'}\,.
\eeq
In fact, the orbit with old orbital parameters $(p,e)$ is shifted along the radial direction.
However, one can recover the usual relation (\ref{rpmdef}) by introducing a new pair of orbital parameters $(p',e')$ with respect to the new coordinate system.
For instance, the new parameters associated with old type II orbits having $p=1+e$ turn out to be
\begin{eqnarray}
p'&=&
\cosh^2r_0\frac{(1+6e+e^2)\cosh(2r_0)-4\sqrt{e}(1+e)\sinh(2r_0)+(1-e)^2}{(1+e)^2\cosh(2r_0)-2\sqrt{e}(1+e)\sinh(2r_0)+(1-e)^2}
\,,\nonumber\\
e'&=&
2\sqrt{e}\frac{2\sqrt{e}\cosh(2r_0)-(1+e)\sinh(2r_0)}{(1+e)^2\cosh(2r_0)-2\sqrt{e}(1+e)\sinh(2r_0)+(1-e)^2}
\,,
\end{eqnarray}
which reduce to $p'=\cosh^2r_0$ and $e'=0$ for $e=0$, corresponding to a circular orbit with radius $r_0$, as for type I orbits in the same limiting case of vanishing eccentricity.


\subsection{Kundt's form and circles}

The cylindrical standard form (\ref{met_godel}) of the metric is related to Kundt's form \cite{kundt}
\beq
\label{met_godel_k}
ds^2=-\left(dT+\frac1{\omega Y}dX\right)^2+\frac1{2\omega^2Y^2}(dX^2+dY^2)+dZ^2\,,
\eeq
by the transformation
\begin{eqnarray}
\label{kundttrasf}
\omega T&=&\sqrt{2}t-2\arctan\left(e^{-2r} \tan \frac{\phi}{2}\right)+\phi\,,\nonumber\\
\sqrt{2}\omega(X+iY)&=&i\frac{1-\tanh r e^{-i\phi}}{1+\tanh r e^{-i\phi}}\,,\qquad
Z=\frac{\sqrt{2}}{\omega}z\,.
\end{eqnarray}
We notice the following Killing vectors
\beq
\Xi_1=\partial_T\,,\qquad
\Xi_2=\partial_X\,,\qquad
\Xi_3=X\partial_X+Y\partial_Y\,,\qquad
\Xi_4=\partial_Z\,,
\eeq
which are combinations of those associated with the cylindrical-like metric (see Eq. (\ref{killcyl}))
\begin{eqnarray}
\Xi_1&=&\frac{\omega}{\sqrt{2}}\partial_t
=\frac{\omega}{\sqrt{2}}\xi_1\,,\nonumber\\
\Xi_2&=&\omega(\partial_t-\sqrt{2}\partial_\phi)+\sqrt{2}\omega\left[\frac{1}{\sqrt{2}}\tanh(r)\cos\phi\,\partial_t-\frac12\sin\phi\,\partial_r-\frac{\cos\phi}{\tanh(2r)}\partial_\phi\right]\nonumber\\
&=&\omega\left(\xi_1-\sqrt{2}\xi_2+\frac{1}{\sqrt{2}}\xi_4\right)\,,\nonumber\\
\Xi_3&=&-\frac{1}{\sqrt{2}}\tanh(r)\sin\phi\,\partial_t-\frac12\cos\phi\,\partial_r+\frac{\sin\phi}{\tanh(2r)}\partial_\phi
=-\frac12\xi_5\,,\nonumber\\
\Xi_4&=&\frac{\omega}{\sqrt{2}}\partial_z
=\frac{\omega}{\sqrt{2}}\xi_3\,.
\end{eqnarray}
When expressed in this coordinate system the azimuthal Killing vector $\xi_2=\partial_\phi$ becomes
\beq
\xi_2=\frac1{\omega}(1-\sqrt{2}\omega Y)\partial_T
-\frac{\sqrt{2}}{4\omega}[1+2\omega^2(X^2-Y^2)]\partial_X
-\sqrt{2}\omega XY\partial_Y
=\Xi_5\,.
\eeq

The timelike geodesic $4$-velocity vector of the metric (\ref{met_godel_k}) is re-expressed as
\begin{eqnarray}
U&=&-\frac{\sqrt{2}}{C}\left(\frac{Y'}{2}-Y\right)\partial_T
+\frac{\sqrt{2}}{C}\omega Y(Y'-Y)\,\partial_X\nonumber\\ 
&&
+\frac{\sqrt{2}}{C}\omega Y(X-X')\,\partial_Y
+\frac{B}{C}\partial_Z\,,
\end{eqnarray}
i.e.,
\beq
U^\flat = -\frac{Y'}{\sqrt{2}C}dT -\frac{1}{\sqrt{2}C\omega}dX+\frac{(X-X')}{\sqrt{2}C\omega Y}dY +\frac{B}{C}   dZ\,,
\eeq
with normalization condition ($U^\alpha U_\alpha=-1$) equivalent to
\beq
\label{normaliz_kundt}
(X-X')^2+(Y-Y')^2=\frac12(Y')^2-B^2-C^2\equiv R^2\,,
\eeq
where $X'$, $Y'$, $C$ and $B$ are constants associated with the above Killing vectors
\begin{eqnarray}
\Xi_1\cdot U&=&-\frac{Y'}{\sqrt{2}C}\,,\qquad
\Xi_2\cdot U=-\frac{1}{\sqrt{2}\omega C}\,,\nonumber\\
\Xi_3\cdot U&=&-\frac{X'}{\sqrt{2}\omega C}\,,\qquad
\Xi_4\cdot U=\frac{B}{C}\,,\nonumber\\
\Xi_5\cdot U&=&\frac1{4\omega^2C}[1-2\sqrt{2}\omega Y'+2\omega^2((X')^2+(Y')^2-R^2)]\,.
\end{eqnarray}

Eq. (\ref{normaliz_kundt}) suggests the following parametric equations for the orbit
\beq
X(\tau)=X'+R\cos\alpha(\tau)\,,\qquad
Y(\tau)=Y'+R\sin\alpha(\tau)\,,\qquad
Z(\tau)=\frac{B}{C}\,\tau\,,
\eeq 
which imply
\beq
\frac{d\alpha(\tau)}{d\tau}=\frac{\sqrt{2}}{C}\omega(Y'+R\sin\alpha(\tau))\,,\qquad
\frac{dT(\tau)}{d\tau}=\frac{\sqrt{2}}{C}\left(\frac{Y'}{2}+R\sin\alpha(\tau)\right)\,,
\eeq
with solution
\beq
\alpha(\tau)=2\arctan\left(\frac{Y'\tan(\Omega\tau)}{\kappa-R\tan(\Omega\tau)}\right)\,,\qquad
\omega T(\tau)= \alpha(\tau)-\frac{Y'}{\sqrt{2}C}\omega \tau\,,
\eeq
where 
\beq
\kappa=\sqrt{(Y')^2-R^2}\,,\qquad
\Omega=\frac{\omega}{\sqrt{2}C}\kappa\,,
\eeq
and initial conditions have been chosen so that $T(0)=0=\alpha(0)$.

Let us now consider planar geodesics corresponding to $B=0$.
Applying the coordinate transformation (\ref{kundttrasf}) to the particle's 4-velocity then yields the following relation between the constants of motion
\beq
E=\frac{Y'}{\omega C}\,,\qquad
L=\frac{C}{2}+\frac{1}{C}\left[\left(\frac{Y'}{2}-\frac1{\sqrt{2}\omega}\right)^2+\frac{(X')^2}{2}-\frac1{4\omega^2}\right]\,,
\eeq
which imply
\begin{eqnarray}
\label{relwithcyl}
\omega X'&=&\pm\left[-\frac12-(\tilde E^2+1)\tilde C^2+2(\tilde E+\tilde L)\tilde C\right]^{1/2}\nonumber\\ 
&\equiv&\pm\sqrt{\tilde E^2+1}\sqrt{(\tilde C-\tilde C_1)(\tilde C_2-\tilde C)}
\,,\nonumber\\ 
\omega Y'&=&\sqrt{2}\tilde E\tilde C
\,,\nonumber\\ 
\omega R&=&\tilde C\sqrt{\tilde E^2-1}
\,,
\end{eqnarray}
with $\tilde C=\omega C>0$ and $\tilde C_1\leq\tilde C\leq\tilde C_2$.
Because $Y'$ is always positive, the centers cannot be located along the $X$-axis.
In contrast, they may lie on the $Y$-axis if either $\tilde C=\tilde C_1$ (with $\omega Y'_1=\sqrt{2}\tilde E\tilde C_1$) or $\tilde C=\tilde C_2$ (with $\omega Y'_2=\sqrt{2}\tilde E\tilde C_2$).
Furthermore, the radius of the orbit is such that $R_1\leq R\leq R_2$.

\begin{figure}
\includegraphics[scale=.35]{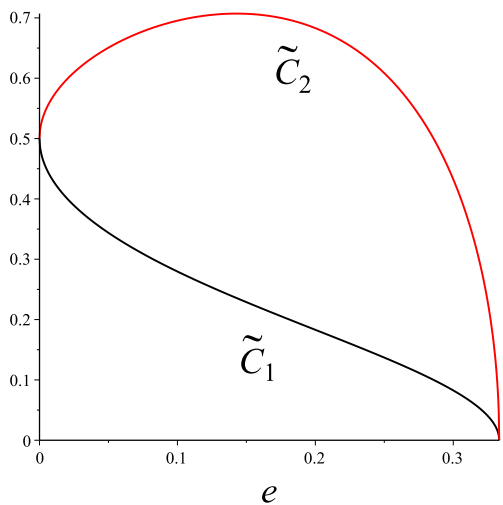}
\caption{
The behavior of the quantities $\tilde C_1$ and $\tilde C_2$ as functions of $e$ is shown for orbits of type II.
}
\label{fig:C12_type2}
\end{figure}

\begin{figure}
\includegraphics[scale=.35]{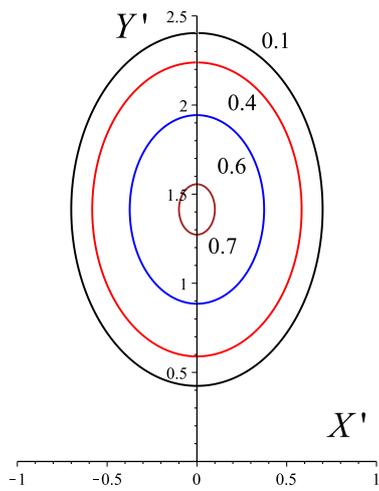}
\caption{
The locus of the centers for type II orbits is shown for selected values of $\tilde C=[0.1,0.4,0.6,0.7]$.
}
\label{fig:centers_type2}
\end{figure}

Let us study the locus of the centers for type II orbits, for fixed values of $\tilde C$.
Substituting $\tilde L=0$ and $\tilde E=\sqrt{{(1+e)}/{(1-3e)}}$ in Eq. (\ref{relwithcyl}) gives
\beq
\tilde C_1=\frac{\sqrt{1-3e}}{2(1-e)}(\sqrt{1+e}-\sqrt{2e})
\,,\qquad
\tilde C_2=\frac{\sqrt{1-3e}}{2(1-e)}(\sqrt{1+e}+\sqrt{2e})
\,.
\eeq
Their behavior as functions of $e$ is shown in Fig. \ref{fig:C12_type2}.
Therefore
\beq
Y'_1=\frac{\sqrt{1+e}}{\sqrt{2}(1-e)}(\sqrt{1+e}-\sqrt{2e})
\,,\qquad
Y'_2=\frac{\sqrt{1+e}}{\sqrt{2}(1-e)}(\sqrt{1+e}+\sqrt{2e})
\,,
\eeq
which become $Y'_1=1/\sqrt{2}=Y'_2$ and $Y'_1=\sqrt{2}-1$, $Y'_2=\sqrt{2}+1$ in the two limiting cases $e\to0$ and $e\to1/3$, respectively.
The locus of the centers is shown in Fig. \ref{fig:centers_type2}.

\begin{figure}
\includegraphics[scale=.35]{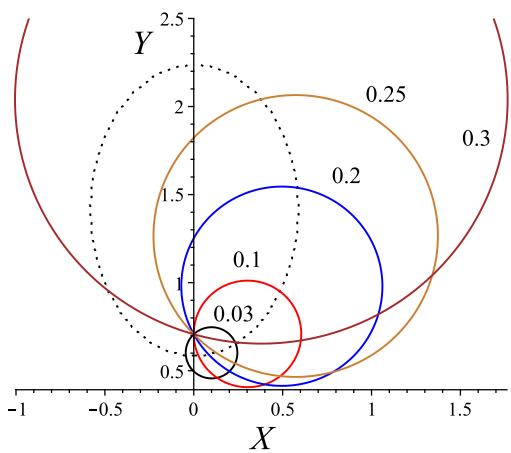}
\caption{
The circles corresponding to type II orbits are shown for $\tilde C=0.4$ and selected values of the eccentricity $e=[0.03,0.1,0.2,0.25,0.3]$.
The dotted curve is the locus of the centers of Fig. \ref{fig:centers_type2}.
Only positive values of $X'$ are taken, for simplicity (see Eq. (\ref{relwithcyl})).
}
\label{fig:circles_type2}
\end{figure}

The radius of the orbit is
\beq
\omega R=2\tilde C\sqrt{\frac{e}{1-3e}}\,,
\eeq
so that its limiting values are
\beq
\omega R_1=\frac{\sqrt{e}}{1-e}(\sqrt{1+e}-\sqrt{2e})
\,,\qquad
\omega R_2=\frac{\sqrt{e}}{1-e}(\sqrt{1+e}+\sqrt{2e})
\,.
\eeq
The orbit intersects the $Y-$axis at $\omega Y=1/\sqrt{2}$ for every value of $\tilde C$ and $e$ and at $\omega Y=2\omega Y'-1/\sqrt{2}$.
Fig. \ref{fig:circles_type2} shows the circles corresponding to type II orbits for a fixed value of $\tilde C$ and different values of the eccentricity $e$.


\section{M\"obius transformations}

A M\"obius transformation $T_{(a,b,c,d)}$ of the complex plane $\mathbb{C}$ has the fractional-linear form 
\beq
\label{mob}
T_{(a,b,c,d)}(z) = \frac{az + b}{cz + d}\,,
\eeq
where $a,b,c,d,z\in  {\mathbb C}$ are complex numbers and $ad-bc\not = 0$.
The set of all M\"obius transformations is a group under composition. The transformations  $S_{(a,b)}(z)=T_{(a,b,0,1)}(z)=az + b$  (setting $d=1$ without loss of generality since one can always take $a\to a/d$, $b\to b/d$) 
form the subgroup of similarities, whereas the transformation $T_{(0,1,1,0)}(z)=1/z$ is termed an inversion. 
Every M\"obius transformation $T_{(a,b,c,d)}$ can be interpreted as a nonunique combination of a similarity and an inversion. 

It is well known that there exists an isomorphism between the Lorentz group SO$(3,1)$ that preserves the orientation of space and the group SL$(2,{\mathbb C})$ of conformal transformations of the 2-dimensional sphere, under which Lorentz transformations correspond to conformal transformations of this sphere (see, e.g., Ref.~\cite{Oblak:2015qia} and references therein), represented by complex $2\times2$ matrices of unit determinant.
The basic geometric property of M\"obius transformations is that they preserve angles between curves and map circles (and straight lines) into circles (and straight lines).

\subsection{Flat spacetime}

Consider the Minkowski metric expressed in cylindrical  coordinates
\beq
\label{flat1}
ds^2= -dt^2 +dr^2+r^2 d\phi^2 +dz^2\,, 
\eeq
simply derived from the standard Cartesian inertial coordinate form
\beq
ds^2= -dt^2 +dx^2+ dy^2 +dz^2\,, 
\eeq
using the coordinate transformation $x=r\cos\phi$, $y=r\sin \phi$ in the $x$-$y$ plane.
Identifying this real plane with the complex plane with the further transformation
$ \zeta =x+i y = r e^{i\phi}$, the inverse transformation is
\beq
\label{map_4}
r =(\zeta\bar \zeta)^{1/2} \,,\qquad \phi=\frac{1}{2i}\ln\left(\frac{\zeta}{\bar\zeta}\right) \,,
\eeq
which leads to the following expression for the metric (\ref{flat1})
\beq
\label{met_tz}
ds^2=-dt^2 +d\zeta d \bar \zeta +dz^2\,,
\eeq
which remains invariant under the following complex  linear  transformation
\beq
\label{linear_comp}
 \zeta' = \sigma\zeta + \beta\,,\qquad \sigma=e^{i\theta}\,,\ \theta\in \mathbb{R}
\eeq
representing a combined rotation and translation of the $x$-$y$ plane 
as long as $\sigma$ is a unit complex number. A pure rotation (set $\beta=0$) corresponds to a counterclockwise rotation $\phi\to\phi+\theta$ by the angle $\theta$ in this plane, or
\beq\label{xyrotation}
  (x,y) \to (x \cos\theta-y \sin\theta, x \sin\theta+y \cos\theta)\,.
\eeq

The Minkowski spacetime timelike geodesics are straight lines representing the world lines of massive test particles of mass $m>0$.  If they are parametrized by the proper time $\tau$, assuming that they pass through the point $(x_0,y_0,z_0)$ of the above inertial coordinate system at $\tau=0$, these take the form
\beq
\label{lines}
t=E \tau\,,\qquad 
x-x_0=p_x \tau \,,\qquad 
y-y_0=p_y \tau\,,\qquad 
z-z_0=p_z \tau\,,
\eeq 
with the normalization condition
\beq
E^2=m^2+p^2\,,\qquad p^2=p_x^2+p_y^2+p_z^2
\eeq
for the associated energy and momentum per unit mass.
The  momentum 1-form (per unit mass) along the geodesic is
\beq
P=-E dt +p_a dx^a\,.
\eeq
Their representation in cylindrical coordinates centered now at $(x_0,y_0,z_0)$ becomes
\beq
t=E \tau\,,\qquad 
r=p_\perp \, \tau \,,\qquad 
\phi=\arctan \left(\frac{p_y}{p_x}\right)\,,\qquad 
z-z_0=p_z \tau\,,
\eeq
where
\beq
p_\perp =\sqrt{p_x^2+p_y^2}\,.
\eeq
We limit our considerations to orbits on a constant $z$ hyperplane by assuming $p_z=0$.
The fact that  $\phi$ is constant along the motion  allows for the parametrization
\beq
p_x=p_\perp \cos\phi\,\qquad 
p_y=p_\perp \sin\phi\,,\qquad 
\phi={\rm const}.
\eeq
or equivalently
\beq
\zeta =r e^{i\phi}=(p_x+ip_y) \, \tau \,,\qquad 
|\zeta|= p_\perp \tau\,.
\eeq
The effect of the transformation (\ref{linear_comp}) is then
\beq
\zeta' =  e^{i\theta} (p_x+ip_y) \, \tau + \beta= (p_x'+ip_y') \, \tau + \beta\,,
\eeq 
rotating the momentum directly analogous to Eq.~(\ref{xyrotation})
\beq
(p_x',p_y') 
=(\cos\theta\, p_x-\sin\theta\, p_y, 
\sin\theta\, p_x+\cos\theta\, p_y)\,.
\eeq

Applying the M\"oebius transformation (\ref{mob}) to the straight lines (\ref{lines}), they are mapped to circles (or again straight lines). 
In fact, the line (\ref{lines}) for $p_z=0$ and $z=z_0$, i.e.,
\beq
p_yx-p_xy-p_yx_0+p_xy_0=0\,,
\eeq 
can be written as
\beq
\label{eqline}
\bar\alpha\zeta-\alpha\bar\zeta-2i(-p_yx_0+p_xy_0)=0\,,
\eeq
where
\beq
\alpha=p_x+ip_y\,.
\eeq
The effect of the map $T_{(0,1,1,0)}(\zeta) =1/\zeta$ to Eq. (\ref{eqline}) is 
\beq
\frac{\bar\alpha}{\zeta}-\frac{\alpha}{\bar\zeta}-2i(-p_yx_0+p_xy_0)=0\,,
\eeq
i.e.,
\beq
\bar\alpha\bar\zeta-\alpha\zeta-2i(-p_yx_0+p_xy_0)\zeta\bar\zeta=0\,,
\eeq
which is the equation of a circle if $-p_yx_0+p_xy_0\not=0$ (or a straight line again if $-p_yx_0+p_xy_0=0$) 
\beq
|\zeta-\zeta_{\rm c}|^2=\rho^2\,,
\eeq
where
\beq
\zeta_{\rm c}=-i\frac{\alpha}{2(-p_yx_0+p_xy_0)}\,,\qquad
\rho=|\zeta_{\rm c}|\,.
\eeq
These circles, however, are no longer geodesics.

\subsection{M\"obius form of the G\"odel metric: taking circles to circles}

Analogous to the Minkowski spacetime introduction of a complex variable $\zeta=r e^{i\phi}$ in the ``polar plane", define 
\beq
\zeta=\tanh r\, e^{i\phi}\,,
\eeq
with inverse map
\beq
r ={\rm arctanh}((\zeta\bar \zeta)^{1/2})=\ln \left( \frac{1+(\zeta\bar \zeta)^{1/2}}{1-(\zeta\bar \zeta)^{1/2}}\right)^{1/2}\,,\qquad 
\phi=\frac{1}{2i}\ln\left(\frac{\zeta}{\bar\zeta}\right)\,.
\eeq
which transforms the cylindrical-like standard metric (\ref{met_godel})  into the form
\begin{eqnarray}
\label{moebiusmet}
ds^2
&=&\frac{2}{\omega^2}\left\{-\left[dt-\frac{i(\zeta d\bar \zeta-\bar \zeta  d\zeta)}{\sqrt{2}(1-\zeta\bar \zeta)}  \right]^2+\frac{d\bar \zeta d\zeta}{(1-\zeta\bar \zeta)^2}+dz^2\right\}\,.
\end{eqnarray}
This derivation needs the differentials
\begin{eqnarray}
&&d\zeta =e^{i\phi} [(1-\tanh^2 r)dr+i \tanh r\, d\phi]\,,
\nonumber\\
&&\bar \zeta d\zeta =\tanh r (1-\tanh^2 r)dr+i \bar\zeta\zeta d\phi \,,
\end{eqnarray}
leading to 
\beq
d\phi=\frac{\bar \zeta d\zeta-\zeta d\bar\zeta}{2i \bar\zeta\zeta}\,, \qquad
dr= \frac{\bar \zeta d\zeta+\zeta d\bar\zeta}{2(\zeta\bar \zeta)^{1/2} (1-\zeta\bar \zeta )}\,.
\eeq

Similarly, the complex map
\beq
\omega T = \sqrt{2}t-i\ln(\zeta)\,,\qquad 
X = \frac1{\zeta}+\bar\zeta\,,\qquad
Y = i\left(\frac1{\zeta}-\bar\zeta\right)\,,\qquad
\omega Z = \sqrt{2}z\,,
\eeq
brings Kundt's form of the metric (\ref{met_godel_k}) into the same M\"obius form (\ref{moebiusmet}).

Successively, the second change of variables
\begin{eqnarray}
t &=& t'+\frac{i\sqrt{2}}{2}\ln\left( \frac{  \bar\alpha_0-\bar\zeta'\beta_0 }{ \alpha_0-\zeta'\bar\beta_0 } \right)\,, \qquad
\zeta =\frac{ \beta_0-\zeta' \bar\alpha_0 }{ -\alpha_0+\zeta'\bar \beta_0 }\,,\nonumber\\
\bar\zeta &=&  \frac{ \bar\beta_0-\bar\zeta'\alpha_0 }{ -\bar\alpha_0+\bar\zeta'\beta_0 } \,,\qquad 
z = z\,,
\end{eqnarray}
leaves the metric formally invariant, i.e.,
\begin{eqnarray}
ds^2
&=&\frac{2}{\omega^2}\left\{-\left[dt'-\frac{i(\zeta' d\bar \zeta'-\bar \zeta'  d\zeta')}{\sqrt{2}(1-\zeta'\bar \zeta')}  \right]^2+\frac{d\bar \zeta' d\zeta'}{(1-\zeta'\bar \zeta')^2}+dz^2\right\}\,,
\end{eqnarray}
as already discussed in Ref.~\cite{Bengtsson}.
Planar timelike geodesics satisfy the equations
\begin{eqnarray}
\dot t&=&\frac{\omega^2{\mathcal E}}{2}+\frac{i(\zeta\dot{\bar\zeta}-\bar\zeta\dot\zeta)}{\sqrt{2}(1-\zeta\bar\zeta)}
\,,\nonumber\\
\dot\zeta\dot{\bar\zeta}&=&\frac{\omega^2}{2}\left(\frac{\omega^2{\mathcal E}^2}{2}-1\right)(1-\zeta\bar\zeta)^2
\,,\nonumber\\
2\zeta\dot{\bar\zeta}^2&=&(1-\zeta\bar\zeta)(i\sqrt{2}\omega^2{\mathcal E}\dot{\bar\zeta}-\ddot{\bar\zeta})\,,
\end{eqnarray}
where a dot denotes derivative with respect to the proper time $\tau$ and $\mathcal E$ is the conserved energy per unit mass of the particle.
Circular geodesics $\zeta=Re^{-i\Omega\tau}$ then exist, with
\beq
R=\frac{\sqrt{\omega^2{\mathcal E}^2+2}-\sqrt{2}\omega{\mathcal E}}{\sqrt{\omega^2{\mathcal E}^2-2}}\,,\qquad
\Omega=\omega\sqrt{\omega^2{\mathcal E}^2+2}\,,
\eeq
and 
\beq
t=\left(\frac{\Omega}{\sqrt{2}}-\frac12\omega^2{\mathcal E}\right)\tau\,.
\eeq

\section{Concluding remarks}

Planar timelike geodesics of the G\"odel spacetime, i.e., their projection on a hyperplane orthogonal to the symmetry axis, have been extensively studied in the literature by using different coordinate systems. 
Among them is the frequently used choice of cylindrical-like coordinates naturally adapted to the spacetime symmetries.
Particle trajectories viewed in these coordinates appear to be elliptic-like and their features can be conveniently studied through an eccentricity-semi-latus rectum $(p,e)$ parametrization familiar (though slightly generalized) from Newtonian mechanics. 
Such deformed ellipses are of three different kinds, depending on the chosen range of orbital parameters, and correspond to closed curves which either go around the origin (type I) or pass through it (type II) or are centered away from it (type III).
Type I orbits may reduce to circles for vanishing eccentricity.
However, since the G\"odel spacetime is homogeneous, then it must look the same from every point.
Therefore, the shape of geodesics about the origin is topologically equivalent to that about any other point chosen as the origin of a new cylindrical-like coordinate system.

The best way to show this feature is to pass to Kundt's form of the G\"odel metric, for which planar geodesics all become circular.
We have established here a one-to-one correspondence between the conserved quantities as well as the parameter space of elliptic-like versus circular geodesics: different kinds of elliptic-like orbits turn out to correspond to circles with different loci of the center as well as different radii depending on the allowed range of orbital parameters $p$ and $e$.
We have also shown how to connect the two equivalent descriptions of particle motion by introducing a pair of complex coordinates in the 2-planes orthogonal to the symmetry axis, which brings the metric into a form which is invariant under M\"obius transformations preserving the symmetries of the orbit, i.e., taking circles to circles.

\end{document}